\documentclass[aps,prl,twocolumn,floatfix,superscriptaddress,10pt,showpacs]{revtex4}

\usepackage{graphicx}
\usepackage{amsmath}
\usepackage{amsfonts}
\usepackage{amssymb}
\usepackage{mathbbol}
\usepackage[colorlinks]{hyperref}
\usepackage[usenames]{color}
\usepackage{multirow}
\usepackage{dcolumn}
\usepackage{eufrak}

\newcommand{\abso}[1]{\left\vert #1 \right\vert}
\newcommand{\ket}[1]{\vert #1 \rangle}
\newcommand{\bra}[1]{\langle #1 \vert}

\newcommand{\scalar}[2]{\left\langle #1 \vert #2 \right\rangle}

\begin{document}

\bibliographystyle{unsrt}

\title{The two classes of low energy spectra in finite carbon nanotubes}

\author{Magdalena Marganska}\thanks{e-mail: magdalena.marganska@physik.uni-r.de}
\affiliation{Institute of Theoretical Physics, Regensburg
University, 93 053 Regensburg, Germany}
\author{Piotr Chudzinski}
\affiliation{Institute of Theoretical Physics, Regensburg
University, 93 053 Regensburg, Germany}
\affiliation{Institute for Theoretical Physics, Leuvenlaan 4,3584 CE Utrecht,The Netherlands}
\author{Milena Grifoni}
\affiliation{Institute of Theoretical Physics, Regensburg
University, 93 053 Regensburg, Germany}

\begin{abstract}
Electrons in carbon nanotubes (CNTs) possess spin and orbital degrees of freedom. The latter is inherited from the bipartite graphene lattice with two inequivalent Dirac points.
The electronic spectra obtained in several transport experiments on CNT quantum dots in parallel magnetic field often show an anticrossing of spectral lines assigned to the opposite Dirac valleys. So far this valley mixing has been attributed to the disorder, with impurity induced scattering. We show that this effect can arise also in ultraclean CNTs of the armchair class and it can be caused solely by the presence of the boundaries. In contrast, in CNTs of the zigzag class it does not occur. These two fundamentally different classes of spectra arise because of different rotational symmetries of the low energy eigenstates in the two types of CNTs.
The magnitude of the level splitting depends in a non-monotonous way on the distance of the involved energy levels from the charge neutrality point.
\end{abstract}
\date{\today}

\maketitle
\begin{figure}[h]
 \includegraphics[width=\columnwidth]{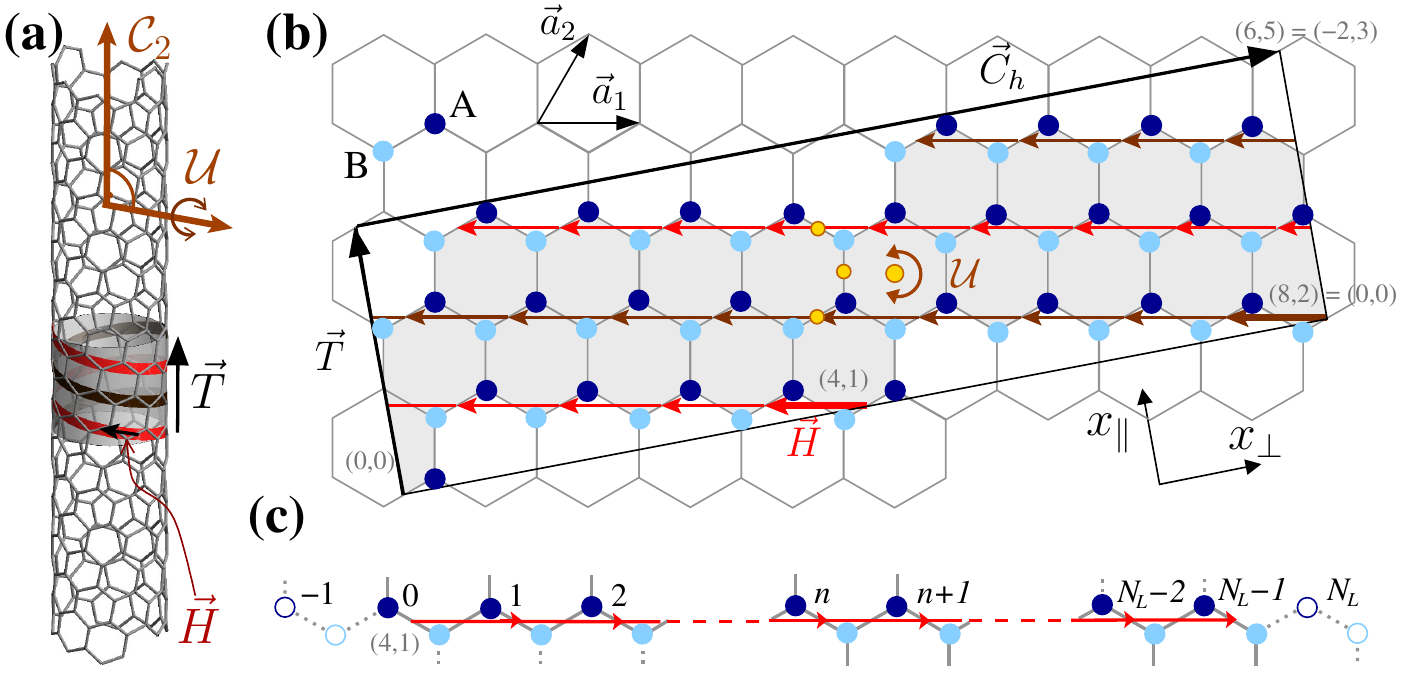}
\caption{\label{fig:symmetries}
Real space structure of a chiral (8,2) nanotube. {\bf (a)} Fragment of a rolled CNT. The translational unit cell with $2N_B$ atoms is marked in grey; fragments of the two helices constituting the CNT are shown in red and brown.  {\bf (b)} Unrolled translational unit cell. A rotation by $\pi$ around one of the $\mathcal{U}$ axes (some choices of $\mathcal{U}$ are marked with yellow dots) maps all atoms onto those from the opposite sublattice. {\bf (c)} An unrolled helical chain from an (8,2) CNT of length $L=N_t|\vec{T}|$. Then $N_L = N_t N_B/\mathfrak{n}$, where $\mathfrak{n}=\mathrm{gcd}(8,2)=2$. Saturated bonds are shown in solid lines, dangling bonds in dotted lines. ``Missing'' atoms are marked by open circles. }
\end{figure}

%
%
{\em Introduction.} 
When the first experiment proving the existence of enhanced spin-orbit (SOI) coupling in CNTs~\cite{kuemmeth:nature2008} was performed, another effect was also observed and later confirmed by other measurements~\cite{jespersen:natphys2011,grove-rasmussen:prl2012,pei:natnano2012,schmid:arxiv2013}. The signature of this phenomenon was an unexpected anticrossing between different valley spectral lines. In the original paper the authors attributed this anticrossing, with its characteristic energy scale $\Delta_{KK'}$, to a $2\vec{K}$ momentum change of an electron, caused by scattering off impurities. The ``disorder term'' or ``valley-mixing'' parameter $\Delta_{KK'}$, of the order of the spin-orbit splitting $\Delta_{SO}$ or larger~\cite{jespersen:natphys2011,schmid:arxiv2013}, is now an accepted part of the CNT Hamiltonian ~\cite{jespersen:natphys2011,laird:arxiv2014}. In a recent study it became even a crucial ingredient, necessary for the emergence of Majorana fermions in CNTs~\cite{sau:prb2013}. 
\\
If disorder is present, it may certainly cause valley mixing~\cite{mccann:prb2005,palyi:prb2010}, with its spectroscopic consequences. However, the experiments revealing the presence of a $KK'$ anticrossing in magnetic fields~\cite{jespersen:natphys2011,grove-rasmussen:prl2012,pei:natnano2012,schmid:arxiv2013} 
have been performed on very clean, most likely disorder-free, nanotube quantum dots. This suggests that 
another mechanism can be responsible for the observed anticrossing. \\
In this work we demonstrate analytically and numerically that this effect can occur also in disorder-free CNTs where it stems from the nanotube's finite size. Its appearance is intimately connected  to the symmetry properties of finite CNTs - it only occurs in CNTs whose low energy subbands in both valleys have the same crystal angular momentum. These CNTs form what we name the armchair or (A) class  - the others belong to the zigzag or (Z) class ~\cite{lunde:prb2005}. In the standard theoretical modelling~\cite{saito:1998,white:prb1993,damjanovic:prb1999,lunde:prb2005,barros:physrep2006} a nanotube is formed by rolling a graphene plane (cf. Fig. \ref{fig:symmetries}(a),(b)), with the circumference of the CNT given by the chiral vector $\vec{C}_h = m_1 \vec{a}_1 + m_2\vec{a}_2$. The pair $(m_1,m_2)$ are the so-called chiral indices of the CNT. Whether a CNT belongs to the (A) or to the (Z) class can be seen immediately from its chiral indices. Let us define $\mathfrak{n}=\mathrm{gcd}(m_1,m_2)$; then if a CNT with chiral indices $(m_1/\mathfrak{n}, m_2/\mathfrak{n})$ is metallic, i.e. $(m_1-m_2)/\mathfrak{n} = 0\vert_{\mathrm{mod}\,3}$, the original $(m_1,m_2)$ CNT is of the (A) class; otherwise it is of the (Z) class.\\
The possibility of breaking the valley degeneracy through different kinds of hard wall boundaries was pointed out in Ref.~\cite{mccann:jpcm2004}. However, the deep relation between CNT's chirality and the nature of the spectrum of a finite tube, as revealed in the present work, was so far unexplored.\\
The requirement of conservation of crystal angular momentum upon reflection off the CNTs' boundaries, together with a generalized parity operation $\mathcal{U}$ (to be defined later), determine the nature of the low energy states at the anticrossing. At the values of magnetic field which counteract $\Delta_{SO}$ for one spin direction the degeneracy between states of opposite $\mathcal{U}$ parity is lifted in (A) class  CNTs, and it is preserved in the (Z) class ones. Hence, the low energy spectra of these two classes of CNTs are qualitatively different.
The strength with which the degeneracy is broken depends on the CNT's length, chirality and the distance of the split energy level from the charge neutrality point (CNP). Because  the $\mathcal{U}$ parity eigenstates are linear combinations of states from different valleys, the splitting $\Delta_{+-}$ in the (A) class CNTs mimics the phenomenological valley mixing parameter $\Delta_{KK'}$. We explore this correspondence at the end of this work.

%
%
{\em Spectrum and symmetries of infinite CNTs.}
The nearest-neighbour tight-binding Hamiltonian of graphene, with one $p_z$ electron per atom, reads
\begin{equation}
\label{eq:hamiltonian-real}
\hat{H} = \sum_{p=A,B} \sum_{\vec{R},\vec{R}'} t_{\vec{R}\vec{R}'} \ket{\vec{R},p}\bra{\vec{R}',-p},
\end{equation}
where $\vec{R},\vec{R}'$ are the lattice vectors and $p$ is the sublattice index (cf. Fig~\ref{fig:symmetries}(b)). For simplicity the spin degree of freedom and spin-orbit interaction effects  will be included later. The hopping integrals $t_{\vec{R}\vec{R}'}$ are all equal in flat graphene, but in CNTs they are modified by the curvature effects \cite{ando:jpsj2000,delvalle:prb2011}. 
In an infinite nanotube we can introduce the sublattice states, defined as $\ket{\vec{k},p} = 1/\sqrt{N} \sum_{\vec{R}} \exp(i\vec{k}\cdot\vec{R})\,\ket{\vec{R},p}$ 
and obtain the Hamiltonian in the reciprocal space,
\begin{equation}
\label{eq:hamiltonian-reciprocal}
 \hat{H} = \sum_{\vec{k}} \underbrace{\sum_{\vec{d}} t_{\vec{R},\vec{R}+\vec{d}} \;e^{i\vec{k}\cdot\vec{d}}}_{\gamma(\vec{k})} \vert \vec{k},A\rangle\langle\vec{k},B\vert + h.c.,
\end{equation}
with $\vec{d}\in\{0,\vec{a}_1,\vec{a}_2 \}$ (cf. Fig~\ref{fig:symmetries}(b)) and $\gamma(\vec{k}) = \vert\gamma(\vec{k})\vert\;\exp(i\eta(\vec{k}))$. 
The Bloch states $\ket{\vec{k}}$ are eigenstates of $\hat{H}$ with energy $\pm\vert\gamma(\vec{k})\vert$, which imposes their form
\begin{equation}
 \ket{\vec{k}}  =  \frac{1}{\sqrt{2}}\left( e^{i\eta(\vec{k})/2} \;\ket{\vec{k},A} \pm e^{-i\eta(\vec{k})/2} \;\ket{\vec{k},B} \right), \label{eq:Bloch}
\end{equation}
where the $(+/-)$ sign applies to the conduction/valence states, respectively. We shall focus on the conduction band.\\
The form of the eigenstates of a nanotube is determined by its symmetries and defined by their associated quantum numbers
\cite{saito:1998,white:prb1993,barros:physrep2006}. Those relevant for us ($\mathcal{C}_{\mathfrak{n}},\mathcal{U},\mathcal{T}(\vec{T}),\mathcal{S}(\alpha,h)=\mathcal{T}(\vec{H})$) are illustrated for the rolled CNT  in Fig.~\ref{fig:symmetries}(a), and their characteristic vectors on the graphene plane  are shown in Fig.~\ref{fig:symmetries}(b). The symmetries, their action on the 3D and 2D lattice and on the Bloch states $\ket{\vec{k}}$, as well as their respective quantum numbers, are listed in Tab. I of the Supplement.\\

\begin{figure}[htb]
\includegraphics[width=0.9\columnwidth]{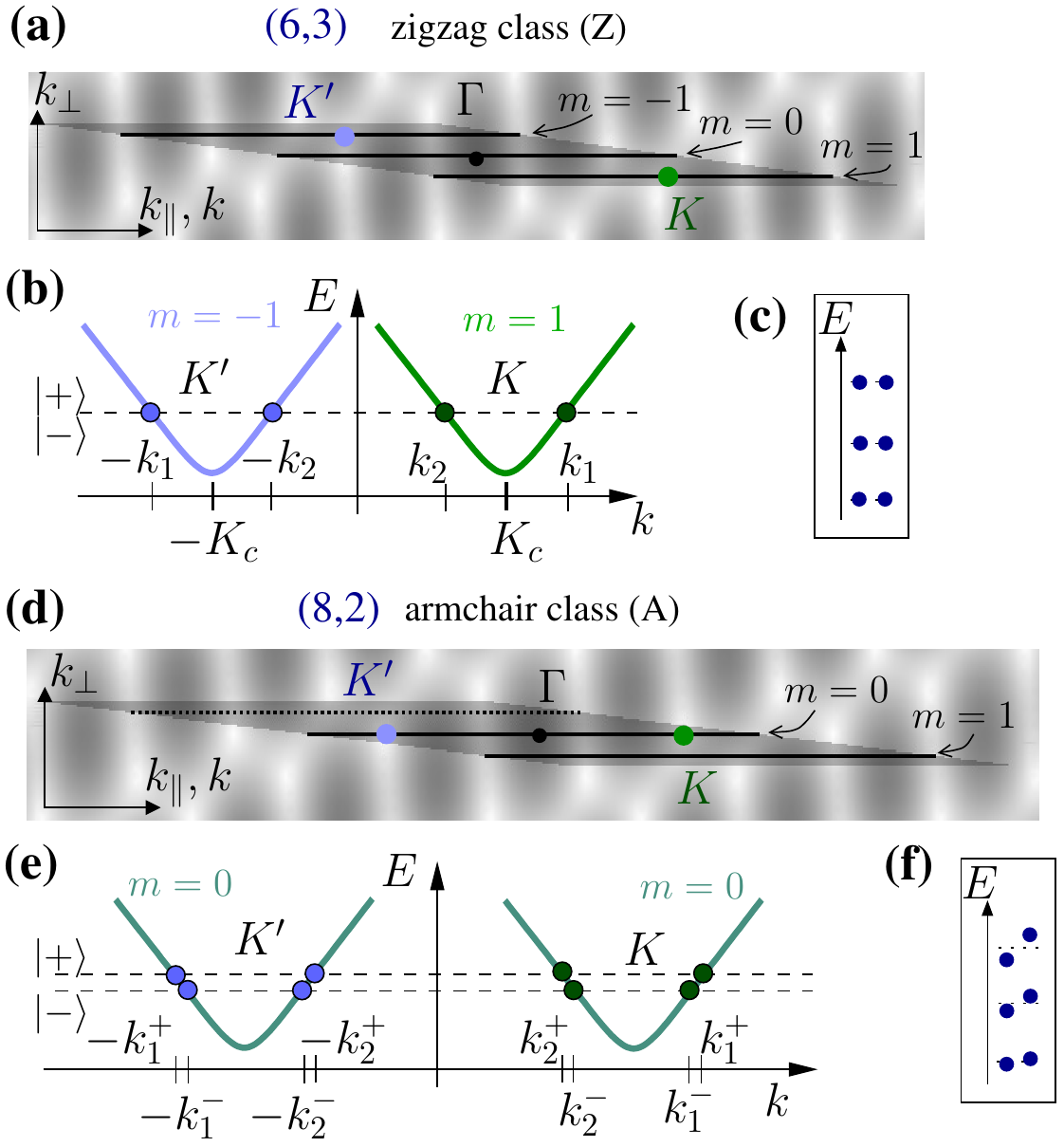}
\caption{\label{fig:finite} Reciprocal space (squeezed in $k_\parallel$ direction) and low energy subbands (without spin) of finite CNTs  in the helical approach. {\bf (a)} Reciprocal unit cell (darker rectangle), $m=const$ subbands and $\Gamma,K,K'$ points of a (Z) class (6,3) CNT. The background is the greyscale plot of graphene's dispersion $\vert\gamma(\vec{k})\vert$. {\bf (b)} Energy subbands. The four degenerate Bloch states of an infinite CNT (points) can be combined into two degenerate $K/K'$ valley or $+/-$ $\mathcal{U}$ parity eigenstates of a finite length CNT. The $K/K'$ valley corresponds to $m=+1/-1$, respectively. {\bf (c)} Part of the doubly degenerate spectrum of a (Z) class CNT. {\bf (d)} As in (a), but for an (A) class (8,2) CNT. The subband $m=-1$ is geometrically equivalent to $m=1$. Both valleys have $m=0$. {\bf (e)} In finite (A) class CNTs the $\mathcal{U}$ parity eigenstates stem from Bloch states with $\vec{k}^+\neq\vec{k}^-$. {\bf (f)} The $\mathcal{U}$ pairs in the resulting spectrum are non-degenerate. }
\end{figure}

The quantum numbers used to identify a Bloch state are most often derived either from the pair of symmetries $(\mathcal{C},\mathcal{T}(\vec{T}))$ (translational scheme) or from $(\mathcal{C}_{\mathfrak{n}},\mathcal{S}(\alpha,h))$ (helical scheme), resulting either in the quantum numbers $(k_\perp,k_\parallel)$ or $(m,k)$, respectively - we will use mostly the $(m,k)$ pair. The rotational symmetries give rise to the quantized transverse quantum numbers - $\mathcal{C}$ to $k_\perp$ and $\mathcal{C}_\mathfrak{n}$ to the crystal angular momentum $m$. Near the CNP the dispersion of graphene forms two inverted cones (valleys), centered around the Dirac points $\vec{K}$ (valley $K$) and $-\vec{K}$ (valley $K'$) in the reciprocal space. The transverse quantization reduces the 2D dispersion to a set of one-dimensional subbands (see Fig.~\ref{fig:finite}(a),(d)). The curvature of the CNT lattice results in a hyperbolic dispersion with minima at shifted Dirac points $\pm \vec{K}_c$ (cf. Fig.~\ref{fig:finite}(b),(e))~\cite{ando:jpsj2000,delvalle:prb2011}. The Dirac momenta $\pm\vec{K}_c$ correspond in the helical representation to $(\pm m,\pm K_c)$, with $m \neq 0$ in CNTs of the (Z) class and $m = 0$ in those of the (A) class~\cite{lunde:prb2005}. Hence, in (A) class CNTs both valleys share the same value of the crystal angular momentum.\\   
Every CNT is also symmetric under a rotation by $\pi$ around an axis $\mathcal{U}$ perpendicular to the $\mathcal{C}$ axis and intersecting either the center of a hexagon or a C-C bond \cite{damjanovic:prb1999} (cf. Fig.~\ref{fig:symmetries}(a),(b)). The action of the $\mathcal{U}$ rotation on atomic orbitals and Bloch states is given by
\begin{equation}                     
\label{eq:U}
\mathcal{U} \ket{\vec{R},p} = \ket{-\vec{R},-p},\quad \mathcal{U}\ket{\vec{k}} = \ket{-\vec{k}}.                                                                                                                                                                                                                                                                                                                                                                                                                                                                                 \end{equation}
In real space $\mathcal{U}$ involves an exchange of sublattices; in reciprocal space, because $\vec{K}' = -\vec{K}$, it involves an exchange of Dirac points (valleys).

%
%
{\em Finite nanotubes.}
Unlike the translational and helical symmetries, neither the transverse symmetries $\mathcal{C},\mathcal{C}_{\mathfrak{n}}$ nor the $\mathcal{U}$ symmetry are broken in finite nanotubes. Therefore we can use them to identify the eigenstates of finite CNTs.
We construct these eigenstates as linear combinations of Bloch states corresponding to the same energy $\vert\gamma(\vec{k})\vert$, imposing on them the symmetry constraints and the boundary conditions. The most general $\mathcal{U}$-symmetric eigenstate is a combination of the four momentum states corresponding to the same $\vert\gamma(\vec{k})\vert$ (see Fig.~\ref{fig:finite}(b),(e)),
\begin{equation}
 \label{eq:eigenstate}
 \ket{\psi}  =  \sum_{i=1,2} |a_i| \left( e^{i\beta_i} \ket{m,k_i} + e^{-i\beta_i}\ket{-m,-k_i}\right).
\end{equation}
We denote the four momenta by $(\pm\vec{k}_1,\pm\vec{k_2})$, with $+\vec{k}$ in the $K$ valley and $-\vec{k}$ in $K'$; the index $1$ stands for the outer branch of the dispersion ($k > K_c$), $2$ for the inner one ($k < K_c$).
In finite nanotubes there is only one choice of $\mathcal{U}$ axis; for one translational cell of an (8,2) CNT it is the one explicitly marked in Fig.~\ref{fig:symmetries}(b). In an arbitrary CNT of length given by $\vec{L}=N_t\vec{T}$ the $\mathcal{U}$ axis is at $(\vec{C}_h+\vec{L}-\vec{H})/2$, with $\vec{H}$ the helical vector (see Tab. I of the Supplement).
The requirement that the energy eigenstates should be also eigenstates of $\mathcal{U}$, $\mathcal{U}\,\ket{\psi^{\pm}}:=\pm\ket{\psi^\pm}$, yields the condition on the phases $\beta_i$:
\begin{equation}
 \vec{k}_i\cdot(\vec{L} - \vec{H}) + 2\pi m n_\alpha/\mathfrak{n} + 2\beta_i = n_i\pi,\quad n_i\in\mathbb{Z}.
 \label{eq:phase-quantization}
\end{equation}
For even (odd) $\mathcal{U}$ eigenstates the integers $n_i$ are even (odd). This choice of $n_i$ ensures the orthogonality between the even and odd $\mathcal{U}$ states, in both CNT classes. Note that {\em a priori} the even and odd states can also have different momenta, $\vec{k}_i^\pm$. The factor $2\pi n_\alpha/\mathfrak{n}$ accounts for the difference between the angular coordinates of the end and the beginning of a helical chain.

\begin{figure}[h]
\includegraphics[width=\columnwidth]{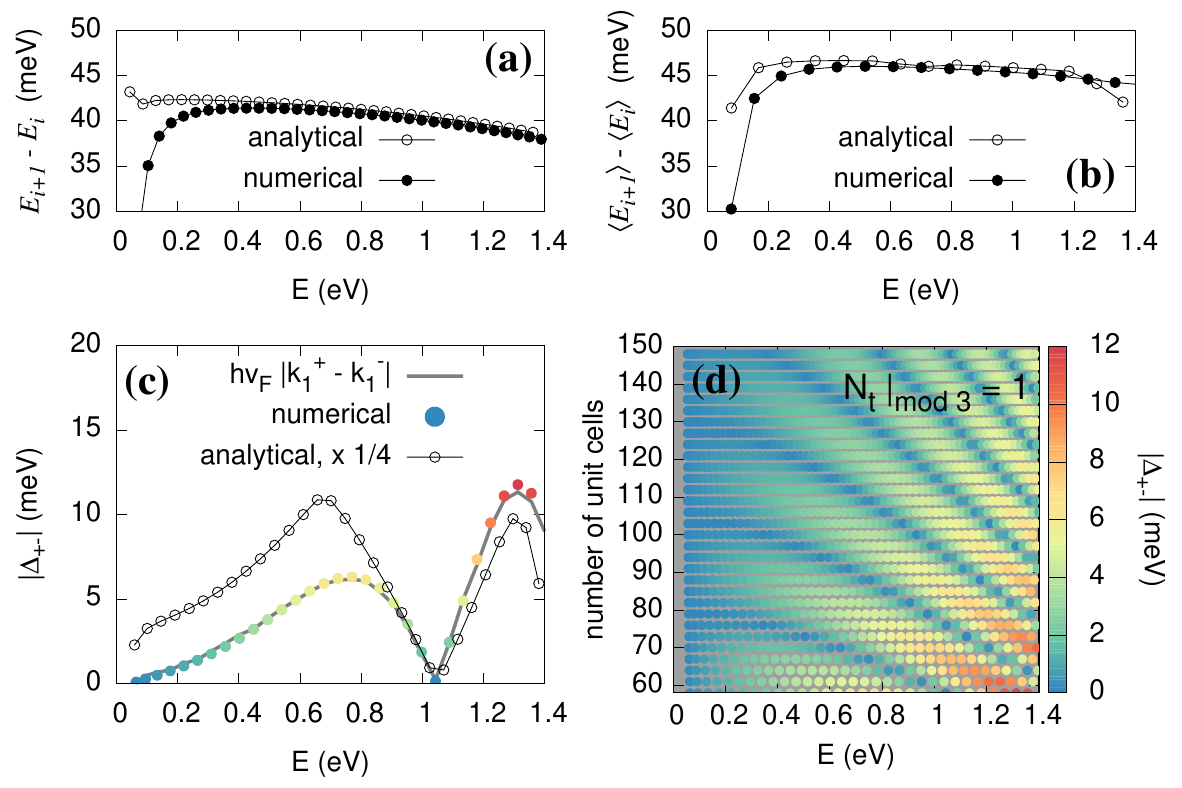}
 \caption{\label{fig:spinless} Spectral properties of finite CNTs without spin-orbit coupling. (a) Spacing between energy levels $E_i$ of a (6,3)x36 CNT of (Z) class.
 (b) Spacing between mean energies of the $\mathcal{U}$ doublets of a (8,2)x58 CNT of (A) class, where $\langle E_i\rangle = (E_i^- + E_i^+)/2$.
  {\bf (c)} The splitting $|\Delta_{+-}|$ vs the mean energy of the shell for an (8,2)x58 nanotube, matching the numerically computed $\hbar v_F |k_i^+ - k_i^-|$. The analytical value of $\Delta_{+-}$ is largely overestimated.
{\bf (d)} The dependence of $|\Delta_{+-}|$ on the length of the CNT and the mean energy of the shell, for $N_t|_{\mathrm{mod}\,3}=1$. The values at the lower edge of the plot, for $N_t = 58$, correspond to the $\Delta_{+-}(E)$ plotted in (c).}
\end{figure}

{\em Boundary conditions.} 
The finite nature of the system is encoded in the boundary conditions. 
We require that the wave functions should vanish at some points, to be specified later, beyond the boundary sites. \\
{\em Zigzag class.} The usual boundary conditions applied in pure zigzag ribbons and CNTs ~\cite{castroneto:rmp2009,delvalle:prb2011} are $\psi_A(\vec{R}_L)=0 \wedge \psi_B(\vec{R}_R)=0$, with $\vec{R}_L$ and $\vec{R}_R$ the first ``missing'' lattice sites beyond the left and right end of the CNT, respectively (see Fig.~\ref{fig:symmetries}(c)). Note that the $\mathcal{U}$ symmetry already implies $\psi_A(\vec{R}_L) = \psi_B(\vec{R}_R = \vec{L}+\vec{C}_h-\vec{H}-\vec{R}_L)$; the finite size of the system in addition sets this common value to 0. When applied to (Z) class CNTs, this yields
\begin{equation}
 \label{eq:quantization-Z}
 \sin\left((\vec{k}_1-\vec{k}_2)\cdot\vec{R}_L' + (\eta_1-\eta_2)/2 \right) = 0,
\end{equation}
with $\vec{R}_L' = \vec{R}_L - (\vec{L}-\vec{H})/2$. 
The left side of this equation defines a ``quantization function'' whose zeroes then yield the momentum eigenvalues, corresponding to discrete longitudinal ($k_\parallel$ or $k$) modes. 
 When the CNT is considered to be a set of helical diatomic chains, the natural choice of the vanishing sites is at $-\vec{H}$ and at $\vec{L}$ from the origin of a chain, as shown in Fig.~\ref{fig:symmetries}(c). 
Let us then set $\vec{R}_L = -\vec{H}$, and at low energies (before the trigonal warping comes into play) Eq. (\ref{eq:quantization-Z}) becomes
\begin{equation}
\kappa_\perp = \kappa_\parallel \cot(\kappa_\parallel (L+h)),
\end{equation}
with $\kappa_\parallel = k_{1\parallel}-K_\parallel^c =K_\parallel^c - k_{2\parallel}$ and $\kappa_\perp = k_{1/2\perp}-K_\perp^c$. This relation, shown here in the translational scheme in which it usually is derived, is the standard momentum quantization condition for pure zigzag CNTs \cite{castroneto:rmp2009,delvalle:prb2011}.\\
The solutions of (\ref{eq:quantization-Z}) determine the momentum pairs $\vec{k}_1(n),\vec{k}_2(n)$, where $\vec{k}_i(n)=\vec{k}_i^+(n)=\vec{k}_i^-(n)$. Out of each pair, with appropriate $\beta_i$'s we can construct two degenerate states; either $\mathcal{U}$ +/- states, or valley $K/K'$ states. The energy spectrum consists of doubly degenerate shells (cf. Fig. \ref{fig:finite}(c) and Section B of the Supplement), numbered by $n$. The level spacing between consecutive $\mathcal{U}$ (or $K/K'$) pairs varies with energy, as shown in Fig.~\ref{fig:spinless}(a). The $\mathcal{U}$ states are then
\begin{equation}
\label{eq:u-finite-z}
\ket{\pm,n}_Z  = \frac{1}{2}\sum_{j=1,2} \abso{a_j} \sum_{\tau=\pm} f^\pm(\tau)\, e^{i\tau\beta_j}\ket{\tau m, \tau k_j(n)},
\end{equation}
where $f^+(\tau)=1$ and $f^-(\tau) = -i\tau$.
The phases $\beta_j$ fulfill the condition (\ref{eq:phase-quantization}), and we have chosen here $\beta_j^- = \beta_j^+ - \pi/2$, with $\beta_j:=\beta_j^+$.
The valley states are linear combinations of $\mathcal{U}$ states,
\begin{equation}
\label{eq:valley-finite-z}
\ket{K/K',n}_Z \equiv \ket{\pm m,n}_Z = \frac{1}{\sqrt{2}} \left( \ket{+,n}_Z \pm i\ket{-,n}_Z\right).
\end{equation}
{\em Armchair class.} In the (A) class CNTs the low energy states belong to the $m=0$ subband. The constraint $\psi_A(\vec{R}_L)=0=\psi_B(\vec{R}_R)$ alone is not sufficient and we must constrain both sublattices at both ends.
It is enough to impose that both $\psi_A$ and $\psi_B$ must vanish on the left end of the CNT - their vanishing on the right follows automatically from the $\mathcal{U}$ symmetry. In the most general case $\psi_A$ and $\psi_B$ can vanish at slightly different positions; a detailed discussion of the choice of $\vec{R}_{L,p}$ can be found in Section B of the Supplement. The set of equations $\psi_A(\vec{R}_{L,A})=0=\psi_B(\vec{R}_{L,B})$ yields a momentum quantization condition
\begin{equation}
\begin{split}
\label{eq:quantization-A}
 0 & = \sin\left( (\eta_1'+\eta_2')/2\right)
\sin\left( (\vec{k}_1-\vec{k_2})\cdot\vec{R}_L'\right) \\
 & \pm \sin\left( (\eta_1'-\eta_2')/2\right)
\sin\left( (\vec{k}_1+\vec{k_2})\cdot\vec{R}_L'\right),
\end{split}
\end{equation}
where $\vec{R}_L' = (\vec{R}_{L,A} + \vec{R}_{L,B} -\vec{L}-\vec{H})/2$ and $\eta_i' = \eta_i - \vec{k}_i\cdot(\vec{R}_{L,A}-\vec{R}_{L,B})$. In contrast to the (Z) class case the quantization function contains now two terms. The first originates from the intra-valley, while the second from the inter-valley backscattering. The finite value of the latter can be traced back to the non-vanishing value of the inter-valley scalar product, $\langle\vec{k}\vert-\vec{k}\rangle\neq 0$. In the (Z) class CNTs $\langle\vec{k}\vert-\vec{k}\rangle=0$; the orthogonality of opposite valley states is protected by the $\mathcal{C}_\mathfrak{n}$ symmetry.\\
The $+/-$ sign in the second line refers to even/odd $\mathcal{U}$ eigenstates, resulting in {\em different quantization} for $\vec{k}^+$ and $\vec{k}^-$. The energies of even and odd states are split by 
\begin{equation}
\label{eq:delta}
\Delta_{+-}(n) = E(\vec{k}^+(n)) - E(\vec{k}^-(n)).
\end{equation}
The (A) class momentum and energy spectrum is sketched in Fig.~\ref{fig:finite}(e)-(f) and discussed further in Section B of the Supplement. The spacing between mean energies of $\mathcal{U}$ pairs varies with energy (cf. Fig.~\ref{fig:spinless}(b)) and so does the $\Delta_{+-}(n)$ (Fig.~\ref{fig:spinless}(c)). The (non-degenerate) $\mathcal{U}$ eigenstates can be written as
\begin{equation}
\label{eq:u-finite-a}
\ket{\pm,n}_A  = \frac{1}{2}\sum_{j=1,2} \abso{a_j^\pm} \sum_{\tau=\pm} e^{i\tau\beta_j^\pm}\ket{0, \tau k_j^\pm(n)}.
\end{equation}
Note that both $\beta_j^\pm$ are now defined by (\ref{eq:phase-quantization}) for different momenta. We may use the same transformation as that which led from (\ref{eq:u-finite-z}) to (\ref{eq:valley-finite-z}) and obtain {\em approximate} valley states,
where the other valley is strongly suppressed, but since $k^+\neq k^-$ the suppression {\em cannot be complete}.\\
For a pure armchair Eq. (\ref{eq:quantization-A}) also holds, but because there $\eta_1 \equiv \pi$ and $\eta_2\equiv 2\pi$, the eigenstates (\ref{eq:eigenstate}) contain only one $\vec{k}_i$, either $\pm\vec{k}_1$ or $\pm\vec{k}_2$. The resulting quantization is that of $2\vec{k}\cdot\vec{R}_L' = n\pi$; it is the only CNT which does behave like a standard quantum box~\cite{rubio:prl1999}.
%
%

{\em Numerical results.}
In order to test the consequences of Eqs. (\ref{eq:quantization-Z}) and (\ref{eq:quantization-A}) we have diagonalized numerically the spinless Hamiltonian (\ref{eq:hamiltonian-real}) for a finite nanotube, with the curvature effects evaluated as in \cite{ando:jpsj2000,delvalle:prb2011} and the values of hopping integrals between $\pi$ and $\sigma$ orbitals set to $V_\pi=-2.66$~eV and $V_\sigma=6.38$~eV~\cite{tomanek:prb1988}.
The resulting low energy spectra of (Z) class CNTs are doubly degenerate, with energy quantization well described by Eq. (\ref{eq:quantization-Z}), see Fig.~\ref{fig:spinless}(a). In (A) class CNTs the $\mathcal{U}$ doublets are split by $\Delta_{+-}$ which depends on the distance from the CNP. The mean shell spacing is well described by 
Eq. (\ref{eq:quantization-A}), but the analytical result overestimates the splitting $\Delta_{+-}$ (cf. Fig.~\ref{fig:spinless}(c) and Section B in the Supplement). 
The Fourier transform of obtained eigenstates shows that the even and odd states indeed have different momenta, $\vec{k}_i^+$ and $\vec{k}_i^-$, whose difference yields $\abso{\Delta_{+-}}\simeq \hbar v_F \abso{k_i^+-k_i^-}$, as shown in Fig.~\ref{fig:spinless}(c). \\
The value of $\Delta_{+-}$ decreases with the length of the CNT (see Fig.~\ref{fig:spinless}(d)), which is natural since it is a finite size effect. It also oscillates with growing amplitude and frequency as the distance from the CNP increases, which is perhaps due to the weaker confinement of the higher energy states. The only systematic experimental investigation of the valley mixing so far~\cite{jespersen:natphys2011} shows that $\Delta_{KK'}$ indeed varies with the gate voltage. \\
In experimental devices the CNT quantum dots are usually defined electrostatically, by the gates. We have modelled (A) and (Z) class CNTs in various soft confining potentials. The division of spectra into two classes, with degenerate or non-degenerate $\mathcal{U}$ doublets, survives, provided that the confinement does not break the $\mathcal{C}_\mathfrak{n}$ symmetry (see Section C of the Supplement). 

\begin{figure}[h]
\includegraphics[width=1.05\columnwidth]{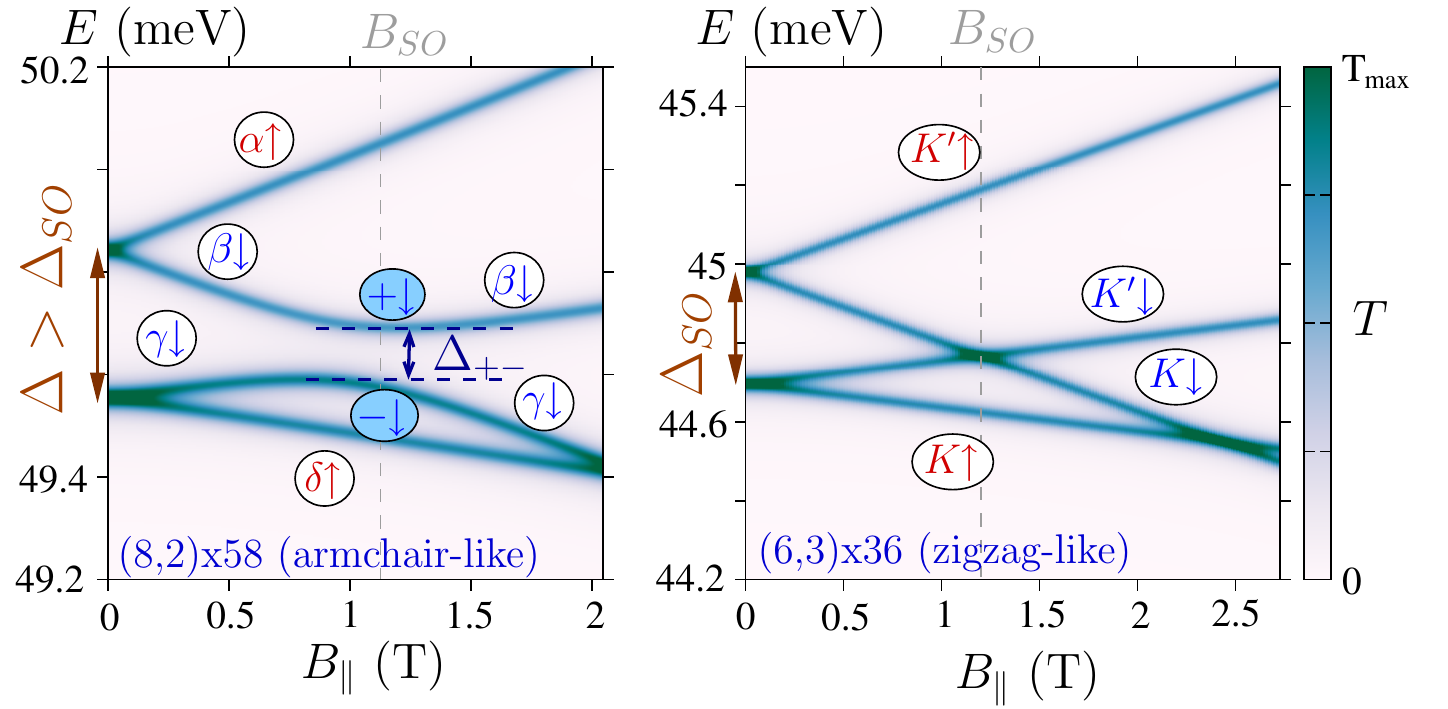}
\caption{\label{fig:magnetic} The transmission of finite, disorder-free, nanotubes in varying axial magnetic field $B_\parallel$. In the chosen energy window only one shell contributes to the transport. The presence of the SOI removes the orbital degeneracy also in (Z) class CNTs and two Kramers pairs split by $\Delta_{SO}$ are present at $B_\parallel =0$. In (A) class CNTs the splitting at zero field $\Delta$ is larger than $\Delta_{SO}$, due to the boundary effects. At $B_\parallel\neq0$ the time-reversal symmetry is broken. The four CNT levels evolve differently:  in (A) class CNTs, when $B_\parallel=B_{SO}$ such that it suppresses $\Delta_{SO}$ for one spin direction, the splitting $\Delta_{+-}$ manifests as an anticrossing. In (Z) class CNTs the valley states remain independent.}
\end{figure}

{\em Inclusion of the spin.} The addition of the electron spin to the problem results, in absence of spin-orbit coupling effects, 
in a doubling of the spectrum - the eigenstates are now $\ket{+/-,\sigma} = \ket{+/-}\otimes\ket{\sigma}$. In (Z) class CNTs the shells $E_n$ are fourfold degenerate and we can express the eigenstates also as $\ket{K/K',\sigma}_Z = \ket{K/K'}_Z\otimes\ket{\sigma}$. In (A) class CNTs each shell contains two doubly degenerate $\mathcal{U}$ states $\ket{+,\sigma}_A$ and $\ket{-,\sigma}_A$ separated by $\Delta_{+-}(n)$.\\
When the SOI is also taken into account, the states $\ket{\vec{k},\sigma}$ and $\ket{-\vec{k},\sigma}$ are no longer degenerate ~\cite{delvalle:prb2011} and the $\mathcal{U}$ symmetry is broken. The time reversal and $\mathcal{C}_\mathfrak{n}$ symmetries are preserved and the spectrum consists of a series of Kramers' doublets. For (Z) class CNTs the shell quantization still follows from (\ref{eq:quantization-Z}), with now spin-dependent $\eta(\vec{k},\sigma)$~\cite{delvalle:prb2011}. The doublets are $\{\ket{K,\sigma},\ket{K',-\sigma}\}$, split by $\Delta_{SO}$. In (A) class CNTs the eigenstates are more complex. The Kramers pairs are $\{ \ket{\alpha,\uparrow},\ket{\beta,\downarrow}\}$ and $\{ \ket{\gamma,\downarrow},\ket{\delta,\uparrow}\}$, where $\ket{\alpha},\ket{\beta},\ket{\gamma},\ket{\delta}$ denote combinations of valley states, determined by the orthogonality requirements $\scalar{\alpha}{\delta}=0=\scalar{\beta}{\gamma}$. This issue will be discussed in a future work, in the following we report on our numerical results.\\
{\em Effects of a parallel magnetic field.} A finite magnetic field breaks the Kramers degeneracy; the four levels in one shell evolve in different ways as the field magnitude is varied (see Fig.~\ref{fig:magnetic}). At some value of the parallel magnetic field the Aharononv-Bohm effect suppresses $\Delta_{SO}$~\cite{delvalle:prb2011}, locally restoring the $\mathcal{U}$ symmetry.  At this field we expect $\Delta_{+-}$ to become apparent in (A) class CNTs. We have tested this hypothesis with numerical calculations of the transmission through finite nanotubes with spin-orbit coupling  in parallel magnetic field using Green's function techniques ~\cite{cuniberti:assp2002,fisher:prb1981}\footnote{The dimensionless parameter $\delta$ measuring the spin-orbit coupling strength \cite{ando:jpsj2000,delvalle:prb2011} was taken to be $\delta = -2\cdot 10^{-4}$.}. This is what was experimentally probed in Ref.~\cite{jespersen:natphys2011,grove-rasmussen:prl2012}. The results for one  CNT of the (A) and one of the (Z) class are shown in Fig.~\ref{fig:magnetic}. 
Initially, at $B_\parallel =0$, the spectra of both CNTs are similar. At the field $B_{SO}$ in (Z) class CNTs the $K/K'$ states with spin $\downarrow$ merely cross, while in the (A) class CNTs the $\Delta_{+-}$ split manifests as an avoided crossing. This effect is present in all shells throughout the low energy spectrum.

%
%
{\em Conclusions.}
The fundamental difference between the (A) and (Z) spectral classes has its roots in the symmetry of the low energy subbands: in (Z) class CNTs the two valleys belong to different representations of the $\mathcal{C}_{\mathfrak{n}}$ group, while in (A) class CNTs they belong to the same representation, with $m=0$. In scattering terms, in finite CNTs of the (A) class the probability amplitude of valley reversal upon reflection off the boundaries does not vanish, $\langle\vec{k}\vert-\vec{k}\rangle\neq 0$. In the (Z) class CNTs their robustness against valley mixing can be crucial in the attempts to use them for quantum computing, relying on the manipulation of spin and valley~\cite{laird:arxiv2014}.
\\
The anticrossing which the $\Delta_{+-}$ introduces is  bound inextricably to the finite size of the system and as such cannot be introduced into the Hamiltonian of an infinite CNT. For the modelling of the energy levels of a CNT quantum dot $\Delta_{+-}$ may be replaced by an effective valley-mixing term, but even that with some caveats - the momenta of $+$ and $-$ states are not the same.

\acknowledgments
We appreciate the discussions with S. Lochner and the financial support from GRK 1570 and SFB 689.

\bibliographystyle{apsrev}

\end{document}